\documentclass{jfm}
\usepackage{commath}
\usepackage{graphicx}
\usepackage{newtxtext}
\usepackage{newtxmath}
\usepackage{natbib}
\usepackage{hyperref}
\usepackage{comment}
\usepackage{color,soul}
\usepackage{amsmath}
\usepackage[normalem]{ulem}
\hypersetup{
	colorlinks = true,
	urlcolor   = blue,
	citecolor  = black,
}

\newcommand{\RomanNumeralCaps}[1]
\linenumbers

\title{Taylor dispersion of bubble swarms rising in quiescent liquid}

\author{Guangyuan Huang\aff{1,2}\footnotemark[3], 
	Hendrik Hessenkemper\aff{1}\footnotemark[3],
	Shiyong Tan\aff{3},
	Rui Ni\aff{3},\\
	Anna-E. Sommer\aff{1},
	Andrew D. Bragg\aff{4}\corresp{\email{andrew.bragg@duke.edu}},
	and Tian Ma\aff{1}\corresp{\email{tian.ma@hzdr.de}}
}

\affiliation{\aff{1} Institute of Fluid Dynamics, Helmholtz-Zentrum Dresden -- Rossendorf, 01328 Dresden, Germany
	\aff{2}Institute of Process Engineering and Environmental Technology, TU Dresden, 01069 Dresden, Germany
	\aff{3} Department of Mechanical Engineering, Johns Hopkins University, MD 21218, USA
	\aff{4}Department of Civil and Environmental Engineering, Duke University, NC 27708, USA}

\begin{document}

\maketitle

\begin{abstract}

We study the dispersion of bubble swarms rising in initially quiescent water using 3D Lagrangian tracking of deformable bubbles and tracer particles in an octagonal bubble column. First, we compare the dispersion inside bubble swarms with that for single-bubble cases and find that the horizontal mean squared displacement (MSD) in the swarm cases exhibits oscillations around the asymptotic scaling predicted for a diffusive regime. This occurs due to wake-induced bubble motion, however, the oscillatory behaviour is heavily damped compared to the single-bubble cases due to the presence of bubble-induced turbulence (BIT) and bubble-bubble interactions in the swarm. The vertical MSD in bubble swarms is nearly an order of magnitude faster than the single-bubble cases, due to the much higher vertical fluctuating bubble velocities in the swarms. We also investigate tracer dispersion in BIT and find that concerning the time to transition away from the ballistic regime, larger bubbles with a higher gas void fraction transition earlier than tracers, consistent with Mathai et al. (\textit{Phys. Rev. Lett.} 121, 054501, 2018). However, for bubble swarms with smaller bubbles and a lower gas void fraction, they transition at the same time. This differing behavior is due to the turbulence being more well-mixed for the larger bubble case, whereas for the smaller bubble case the tracer dispersion is highly dependent on the wake fluctuations generated by the oscillating motion of nearby bubbles.

\end{abstract}


\footnotetext[3]{G. H. and H. H. contributed equally and are joint first authors.} 

\section{Introduction}\label{sec: introduction}

Turbulent dispersion is relevant for understanding many natural and engineered flows, e.g., the spreading of plankton in aquatic ecosystems, the transmission of hazardous aerosols in the atmosphere, and the mixing of components in process engineering, to name a few \citep{2009_Toschi,2014_Bourgoin,2022_Brandt}. When the particles cannot be approximated as tracers, their inertia and the buoyancy forces to which they are subject greatly complicate the dispersion under the influence of the turbulence. One important case is that of bubbles, and understanding the dispersion of bubbles in turbulent flows is crucial for practical multiphase systems and provides fundamental insights into the Lagrangian dynamics of buoyant particles in general \citep{2018_Lohse,2020_Mathai}.

A simple metric for quantifying particle dispersion is the mean squared displacement (MSD) of a particle along its trajectory. In the present work, we denote the particle position by the vector $\boldsymbol{x}(t)$ with Cartesian components $x(t), y(t), z(t)$. For the component $x(t)$, the MSD is
\begin{equation} \label{eqn: MSD1}
	\sigma^2(\Delta_{\tau}x) \equiv \langle (\Delta_{\tau}x - \langle \Delta_{\tau}x\rangle)^2\rangle ,
\end{equation}
where the displacement is $\Delta_{\tau}x \equiv x(t_0+\tau)-x(t_0)$, $\tau$ is the time lag, $t_0$ is the initial time, and $\langle\cdot\rangle$ denotes an average over particles (assuming a homogeneous flow so that the initial position does not matter). Note that throughout this paper, the notation $\sigma(\xi)$ is used to denote the standard deviation of the arbitrary variable $\xi$, such that $\sigma(\Delta_{\tau}x)$ is the standard deviation of $\Delta_{\tau}x$.

For tracer particles in isotropic turbulence, \cite{1922_Taylor} showed that
\begin{equation} \label{eqn: MSD2}
	\sigma^2(\Delta_{\tau}x) \sim
	\begin{cases} 
		\sigma^2(u_l) \tau^2, & \text{for } \tau \ll T_{L,u_l} \text{(ballistic)}\\ 
		2\sigma^2(u_l) T_{L,u_l} \tau, & \text{for } \tau \gg T_{L,u_l} \text{(diffusive)},
	\end{cases}
\end{equation}
where $T_{L,u_l} \equiv \int_{0}^{\infty}R_{u_lu_l}(\tau) \, d\tau$ is the fluid Lagrangian integral time scale, $R_{u_lu_l}(\tau) \equiv \langle u_l(t_0)u_l(t_0+\tau) \rangle/\sigma(u_l)^2$ is the fluid velocity autocorrelation function, $\boldsymbol{u}_l$ is the liquid velocity vector with Cartesian components $u_l,v_l,w_l$. We use the subscript ‘\textit{l}’ for the liquid velocity while the subscript ‘\textit{b}’ is used for the bubble velocity, such that the bubble velocity vector is $\boldsymbol{u}_b$ with Cartesian components $u_b,v_b,w_b$, and the bubble Lagrangian integral time scale is $T_{L,u_b}$.

In contrast to the case of small particles \citep{2022_Brandt}, studies on the turbulent dispersion of finite-sized bubbles are scarce. The numerical study of finite-size bubble dispersion for dense swarms is challenging, not only because such bubbles require interface-resolved direct numerical simulations \citep{2021_Innocenti}, but also because rising bubble dispersion requires long simulation times and large domains. Experiments on the dispersion of millimeter size bubbles are also rare, with \cite{2018_Mathai} being the only published study on this, to the best of the authors’ knowledge. They studied the dispersion of 1.8 mm air bubbles in nearly homogeneous isotropic turbulence (HIT). The gas void fraction is very low in their cases ($\alpha\approx0.05\%$), so there are negligible bubble-bubble interactions and negligible bubble-induced turbulence (BIT). Their results revealed that two mechanisms play a crucial role in the bubble dispersion: the bubble wake dynamics and the crossing-trajectory effect. At short times in the ``ballistic regime'', large wake-induced fluctuations in the bubble velocities causes their MSD to grow much faster than that for tracer particles. At longer times, the MSD approaches the diffusive regime that also occurs for tracer particles. However, because the bubbles can drift through turbulent eddies due to the buoyancy force they experience (``crossing-trajectory effect''), the diffusion coefficient is smaller for the bubbles than the tracers. The time at which the MSD transitions from ballistic-to-diffusive growth is much sooner for the bubbles than the tracers, and they showed that for the bubbles the transition time is related to the intrinsic frequency of vortex shedding from the rising bubble.

While \cite{2018_Mathai} studied the dispersion of small air bubbles with low gas void fraction where the bubbles are relatively isolated, the present study aims to explore how the dispersion of a swarm of bubbles differs from that of a single isolated bubble. Moreover, while the turbulence in their flow was generated using an active grid with negligible BIT, we consider a flow in which all of the turbulence is BIT. To achieve this we consider much larger bubbles with a much higher gas void fraction, leading to strong bubble-bubble interactions and BIT. Such conditions are found both in industrial bubble columns for mixing and in natural bubble plumes.

We seek to answer several specific fundamental questions about the dispersion of bubbles in BIT: How does the dispersion of bubbles rising in swarms differ from that of a single bubble in different directions? How does the dispersion of bubble swarms compare to that of tracer particles? To what extent does the bubble size influence the results? To address these questions, we use experiments to track $O(10^5)$ deformable bubbles using our recently developed 3D Lagrangian bubble tracking (3D-LBT) tool \citep{2024_Hessenkemper}, which can handle overlapping bubbles (crucial for the higher gas void fraction in this study). We also use the open-source 3D Lagrangian particle tracking (3D-LPT) code \citep{2020_Tan} to characterize the liquid phase for comparison.

\section{Experimental method}\label{sec: Experimental method}

\begin{figure}	
	\begin{minipage}[b]{1.0\linewidth}
		\begin{minipage}[b]{0.5\linewidth}
			\centering
			\makebox[1.5em][l]{\raisebox{-\height}{(\textit{a})}}%
			\raisebox{-\height}{\includegraphics[height=4.0cm]{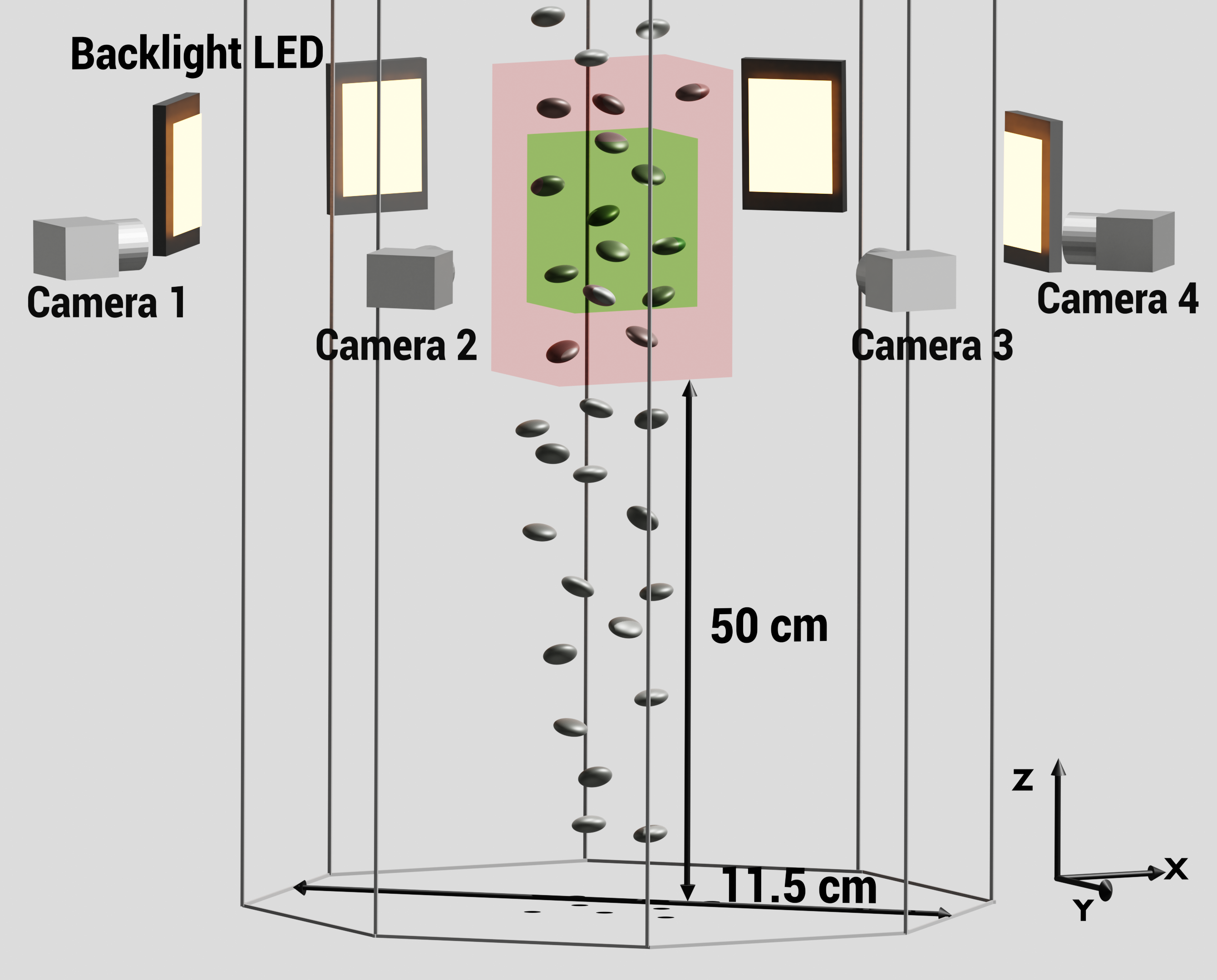}}
		\end{minipage}
		\begin{minipage}[b]{0.5\linewidth}
			\centering
			\makebox[1.5em][l]{\raisebox{-\height}{(\textit{b})}}%
			\raisebox{-\height}{\includegraphics[height=4.0cm]{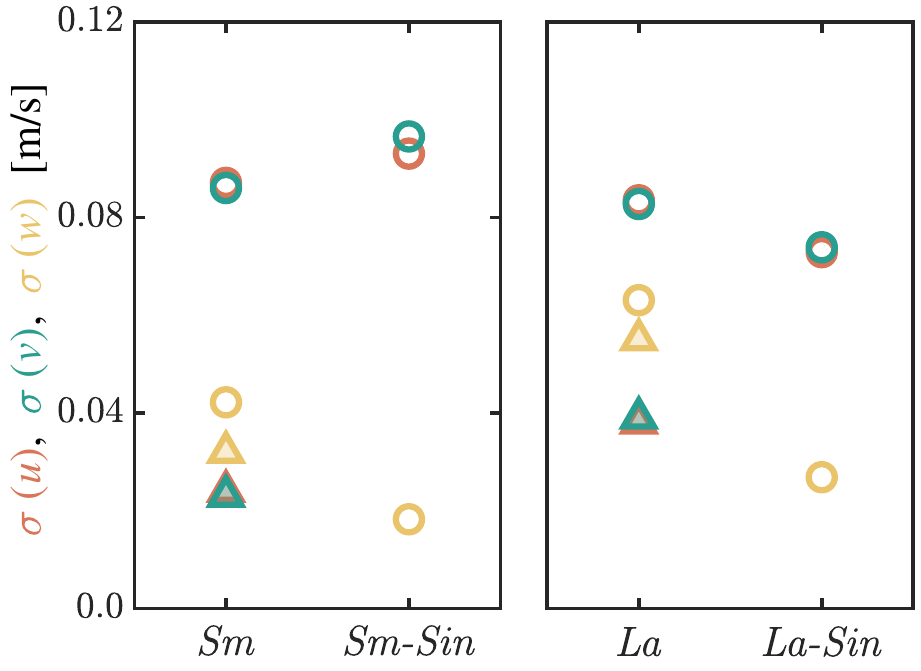}}
		\end{minipage}
	\end{minipage}	
	\caption{(\textit{a}) Schematic of the experimental setup. The red box indicates the bubble tracking measurement section, while the green box is the ROI for 3D-LPT. (Note that in the actual experiment, the number of bubbles in the red box is $O(10^2)$.) (\textit{b}) Standard deviation of the velocity for bubbles (circles) and tracers (triangles), with the three components represented by different colors. Note that the velocity statistics of the liquid are only given for the two swarm cases: \textit{Sm} and \textit{La}.} \label{fig: exp setup}
\end{figure}

The experiments are conducted in an octagonal bubble column that has been optimized for multi-view imaging and 3D tracking (see figure \ref{fig: exp setup}\textit{a}), and is filled with tap water to a height of 900 mm. Air bubbles are injected through needle spargers at the bottom, and two bubble sizes (3.5 mm and 4.4 mm) are considered in the present work by using spargers with different inner diameters. The gas injection rate of all eight spargers is adjusted to produce nearly homogeneous, monodisperse bubbly flows in the column core. For comparison, we also perform single-bubble experiments with the same corresponding bubble size. Only one sparger is operated at a very low flow rate to produce bubbles that are suffciently separated for them to be considered individual (i.e. they experience negligible flow disturbance due to the other bubbles and bubble-bubble interactions do not occur). In total, we investigate four different cases denoted as \textit{Sm}, \textit{La}, \textit{Sm-Sin}, and \textit{La-Sin} (see supplementary movies for each individual case). Here, ‘\textit{Sm/La}’ represents smaller/larger bubbles, and ‘\textit{-Sin}’ stands for the corresponding single-bubble cases. Some key parameters of the two bubble swarm cases are summarized in table \ref{tab: bubble para}. The measurement set-up and data acquisition are similar to those in our previous work \citep{2025_Ma}, so only the key aspects for 3D-LBT and 3D-LPT are stated in what follows.

\begin{table}
	\begin{center}
		\def~{\hphantom{0}}
		\setlength{\tabcolsep}{8pt}
		\begin{tabular}{ccccccc}
			
			&\textsl{$\alpha$}&\textsl{$d_b\,(mm)$}&\textsl{$Re_b$}&\textsl{$T_{L,u_l} (s)$}&\textsl{$T_{L,w_l} (s)$}&\textsl{$Re_{\lambda}$}\\
			\hline
			\textit{Sm} &0.52\%&3.5&832&0.20&0.26&57\\
			\textit{La}  &1.20\%&4.4&920&0.11&0.33&91\\
			
		\end{tabular}
\caption{Selected parameters of the two bubble swarm cases. Here, $\alpha$ the gas void fraction, $d_b$ the bubble diameter, $Re_b$ the bubble Reynolds number based on the bubble to fluid relative velocity obtained from the experiment. For liquid properties, $T_{L,u_l}$ and $T_{L,w_l}$ are the horizontal and vertical Lagrangian integral timescales computed using the associated autocorrelation functions, and $Re_{\lambda}$ the Taylor Reynolds number.} \label{tab: bubble para}
	\end{center}
\end{table}

Four high-speed cameras arranged in a linear configuration are employed to capture the 3D dispersion of bubbles and tracers, with the flow being back-illuminated by LEDs for shadowgraph imaging. The Region of Interest (ROI) for bubble tracking is $5\times5\times15$ cm$^3$ with a recording rate of 500 f.p.s.. Bubbles are identified and tracked using our in-house 3D-LBT tool \citep{2024_Hessenkemper}, specifically designed for tracking relatively dense, deformable bubbles that partially overlap with others in the images. For each case, we collect over $200,000$ bubble tracks to ensure statistical convergence. To capture the 3D flow information, we seed the water with 100 µm polyamide particles as tracers. Since these tracer particles are significantly smaller than bubbles, we perform separate measurements for tracers with a different ROI of $4.5\times4.5\times9.0$ cm$^3$ at a frame rate of 2500 f.p.s., ensuring sufficient spatio-temporal resolution for the flow. The tracers are tracked in 3D using OpenLPT \citep{2020_Tan} -- an open-source version of the Shake-The-Box approach \citep{2016_Schanz} that can handle high particle image densities. 

Figure \ref{fig: exp setup}(\textit{b}) presents all components of the standard deviation of the bubble fluctuating velocities for all four cases as well as the standard deviation of the liquid velocities for the two swarm cases. Consistent with previous studies \citep{2013_Lu}, the liquid velocity fluctuations in both \textit{Sm} and \textit{La} are anisotropic, with larger fluctuations in the vertical direction (corresponding to the direction of the mean bubble motion). In contrast, the bubble velocity fluctuations in all four cases are larger in the horizontal direction, which we will discuss in detail in \S\,\ref{sec: 3.1}.

\section{Results} \label{sec: Results}

\subsection{Dispersion of bubble swarms versus single bubbles}\label{sec: 3.1}

Although the asymptotic results in equation \eqref{eqn: MSD2} were derived for tracer particles, they are based on the kinematic equation $\dot{\boldsymbol{x}}(t)\equiv \boldsymbol{u}_l(t)$ and make no assumption about the dynamics of the particles or flow. As such, exactly the same predictions apply to the case of bubbles dispersing in turbulence, but with $\sigma(u_l)$ replaced by $\sigma(u_b)$, and $T_{L,u_l}$ replaced by $T_{L,u_b}$.

\begin{figure}	
	\begin{minipage}[b]{1.0\linewidth}
		\begin{minipage}[b]{0.5\linewidth}
			\centering
			\makebox[1.3em][l]{\raisebox{-\height}{(\textit{a})}}%
			\raisebox{-\height}{\includegraphics[height=4.7cm]{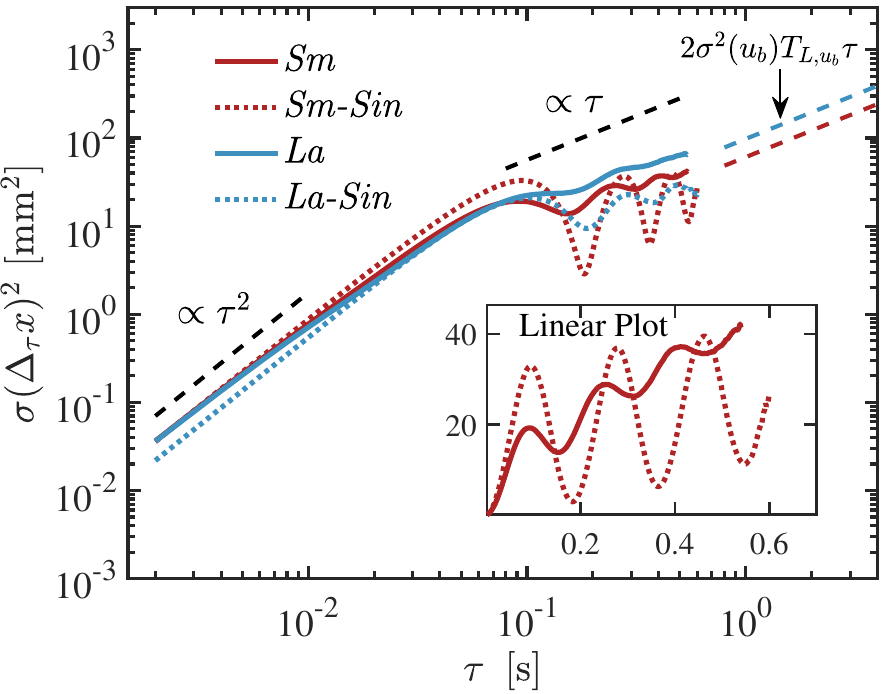}}
		\end{minipage}
		\begin{minipage}[b]{0.5\linewidth}
			\centering
			\makebox[1.3em][l]{\raisebox{-\height}{(\textit{b})}}%
			\raisebox{-\height}{\includegraphics[height=4.7cm]{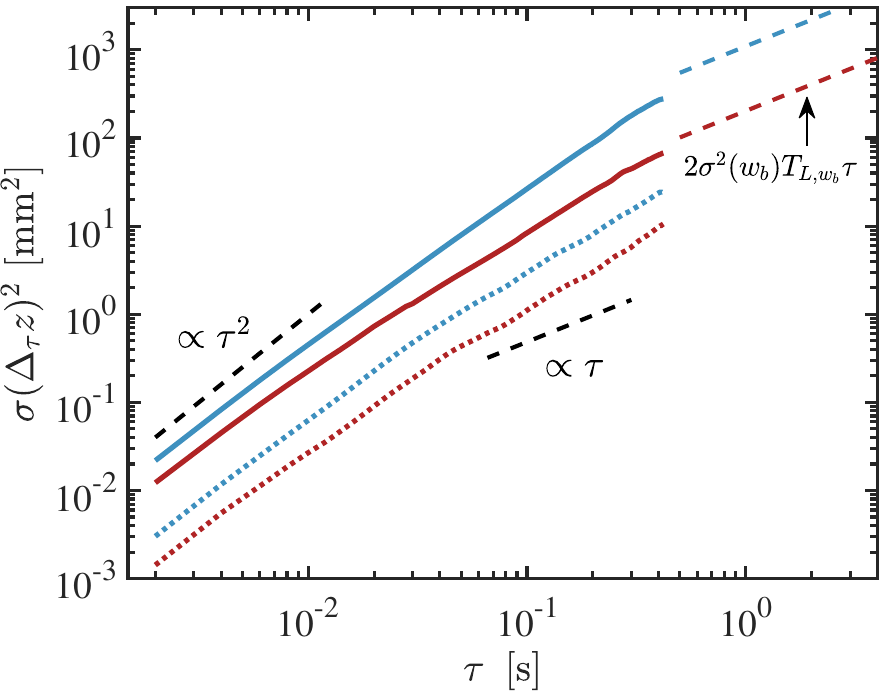}}
		\end{minipage}
	\end{minipage}	
	\begin{minipage}[b]{1.0\linewidth}
		\vspace{2mm}
		\begin{minipage}[b]{0.5\linewidth}
			\centering
			\makebox[1.3em][l]{\raisebox{-\height}{(\textit{c})}}%
			\raisebox{-\height}{\includegraphics[height=4.7cm]{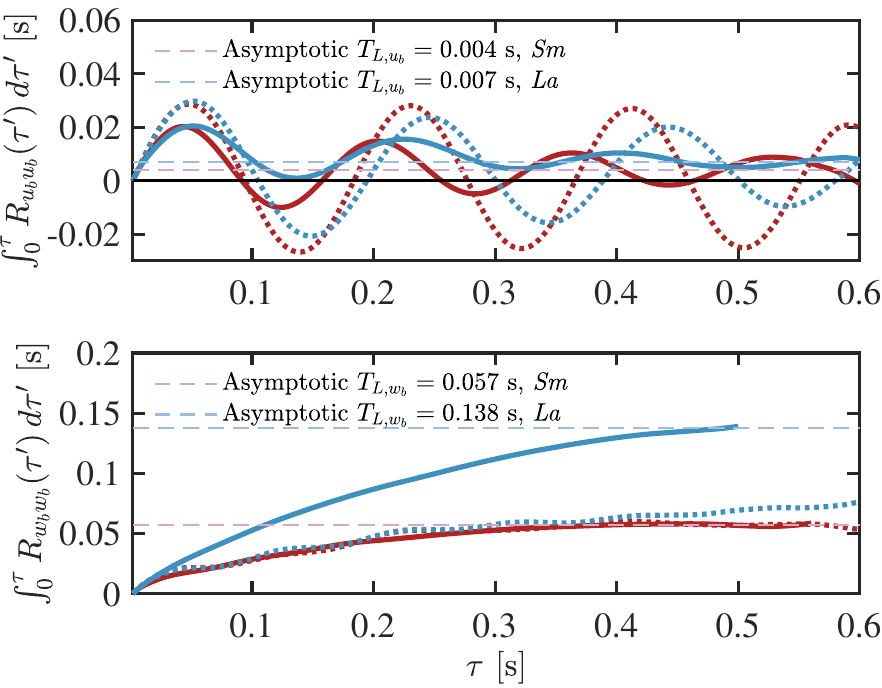}}
		\end{minipage}
		\begin{minipage}[b]{0.5\linewidth}
			\centering
			\makebox[1.3em][l]{\raisebox{-\height}{(\textit{d})}}%
			\raisebox{-\height}{\includegraphics[height=4.7cm]{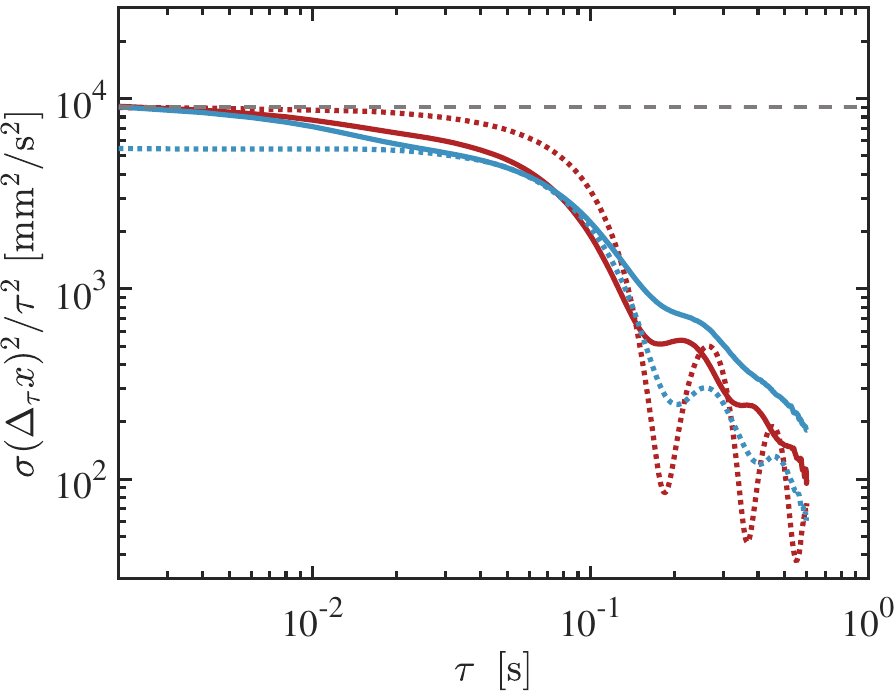}}
		\end{minipage}
	\end{minipage}	 
	
	\caption{MSD for single bubbles (dashed lines) and bubble swarms (solid lines) in the (\textit{a}) horizontal and (\textit{b}) vertical directions. Dashed lines: Asymptotic MSD for longtime diffusive behavior. (\textit{c}) Integral of Lagrangian autocorrelation function of bubble velocity in the horizontal and vertical components, respectively. $T_{L,u_b}$ and $T_{L,w_b}$ are estimated from the asymptotic value of the curve (dashed lines). (\textit{d}) Horizontal MSD compensated by $\tau^2$ to emphasize the deviation from the ballistic regime with increasing time. }
    \label{fig: MSD1}
\end{figure}

From the 3D bubble tracks, we compute the horizontal (figure \ref{fig: MSD1}\textit{a}) and vertical (figure \ref{fig: MSD1}\textit{b}) components of the MSD for the bubbles as a function of time lag $\tau$ for all four cases.
%
For small $\tau$, the MSD of both bubble swarms (\textit{Sm/La}) and single bubbles (\textit{Sm-Sin/La-Sin}) exhibit a $\tau^2$ growth in both directions, as bubbles essentially follow a ballistic motion. This ballistic motion occurs because for $\tau$ much smaller than the correlation timescale of the fluctuating bubble velocities, the bubble velocities are constant to leading order in $\tau$. For the vertical direction, we find that the MSD for bubble swarms is much larger than that for single bubbles (figure \ref{fig: MSD1}\textit{b}). This difference is due to the BIT experienced by the bubbles dispersing in swarms, and the anisotropy of this BIT. It is known that rising bubble swarms generate larger vertical than horizontal fluctuating velocities in the flow \citep{2010_Riboux}, with ratios $\sigma(w_l)/\sigma(u_l)$ of 1.33 and 1.45 for the \textit{Sm} and \textit{La} cases, respectively. Related to this, $\sigma(w_b)$ is much larger for bubble swarms than single-bubbles due to the BIT, and $\sigma(w_b)$ is about three times greater for the swarm cases than the corresponding single-bubble cases (see figure \ref{fig: exp setup}\textit{b}). Since the vertical MSD is proportional to $\sigma^2(w_b)$, this results in an order of magnitude increase in the vertical MSD of bubble swarms compared with single bubbles in the ballistic regime. In contrast, the horizontal component of the MSD does not show such significant differences, since the variations in $\sigma(u_b)$ (or $\sigma(v_b)$) between the single-bubble and swarm cases are relatively small. This is because the horizontal fluctuations of the bubble velocity are dominated by wake-induced instabilities which are common to both the single and swarm cases. 

As $\tau$ increases, the MSD gradually deviates from the ballistic behaviour. The most striking behavior is for the horizontal direction (figure \ref{fig: MSD1}\textit{a}); for the two single-bubble cases the MSD oscillates strongly while for the swarm cases, although the oscillations in the MSD are present (the linear inset plot shows this more clearly) they are strongly damped. The oscillations in the MSD for the \textit{Sm-Sin} and \textit{La-Sin} cases occurs due to the wake instability that leads to an oscillatory rising path for the bubbles \citep{2025_Legendre}. For the two swarm cases, the MSD oscillations are much weaker because the chaotic effects of BIT and bubble-bubble interactions suppresses the systematic oscillatory motion. The oscillations are in fact almost unnoticeable for the \textit{La} case, where higher BIT and stronger bubble-bubble interactions further suppress the oscillations compared to the \textit{Sm} case. Interestingly, the damped MSD oscillations share similarities with that observed for active self-propelled particles moving at the air-water interface in a Petri dish \citep{2020_Bourgoin}. In that case the oscillatory MSD of a single active particle is damped with increasing particle number due to the particle interactions, paving the way for a diffusive MSD behaviour. Although the origin of the oscillations in the MSD is different for the isolated active particle studied in \cite{2020_Bourgoin} (where it is caused by reorientation of the particle trajectory upon hitting the dish boundary) and for an isolated bubble in the present study (where it is caused by wake instabilites), the attenuation of the oscillations with increasing particle/bubble number in both situations reflects the randomizing effect of the collective motion on the individual particle/bubble motion.

At the largest time for which the MSD is computed $\tau=\tau_{max}$, the behavior does not actually follow a diffusive behavior $\propto \tau$. The data in figure \ref{fig: MSD1}(\textit{c}) indicates that this is because the integral of the bubble velocity autocorrelation over the time span $[0,\tau]$ does not converge to a constant as $\tau\to\tau_{max}$ (although it almost converges for the vertical components). For the swarm cases the integral oscillates about a mean value with amplitude that decreases with increasing $\tau$. We can estimate the integral timescale from the asymptotic value of the curve at large $\tau$, and using these we show in figure \ref{fig: MSD1}(\textit{a,b}) the diffusive behavior (e.g. $\sigma^2(\Delta_{\tau}x)=2\sigma^2(u_b) T_{L,u_b}\tau$) that would be expected for time larger than $\tau_{max}$ for the swarm cases. These indicate that the MSD is in fact approaching the expected diffusive behavior. For the single bubble cases, the amplitude of the oscillations of the integrals decreases very slowly with time. Assuming the integrals would eventually converge, this indicates that extremely large times are required to observe diffusive growth for the single-bubble cases.

We note that the transition away from the ballistic regime occurs earlier for the swarm cases, which can be better seen in the compensated horizontal MSD in figure \ref{fig: MSD1}(\textit{d}). In the swarm cases, bubbles maintain their initial velocity for less time than single bubbles due to the randomizing effects of BIT and hydrodynamic interactions with other bubbles in their vicinity. 
\begin{figure}	
	\begin{minipage}[b]{1.0\linewidth}
		\begin{minipage}[b]{0.5\linewidth}
			\centering
			\makebox[1.2em][l]{\raisebox{-\height}{(\textit{a})}}%
			\raisebox{-\height}{\includegraphics[height=5.3cm]{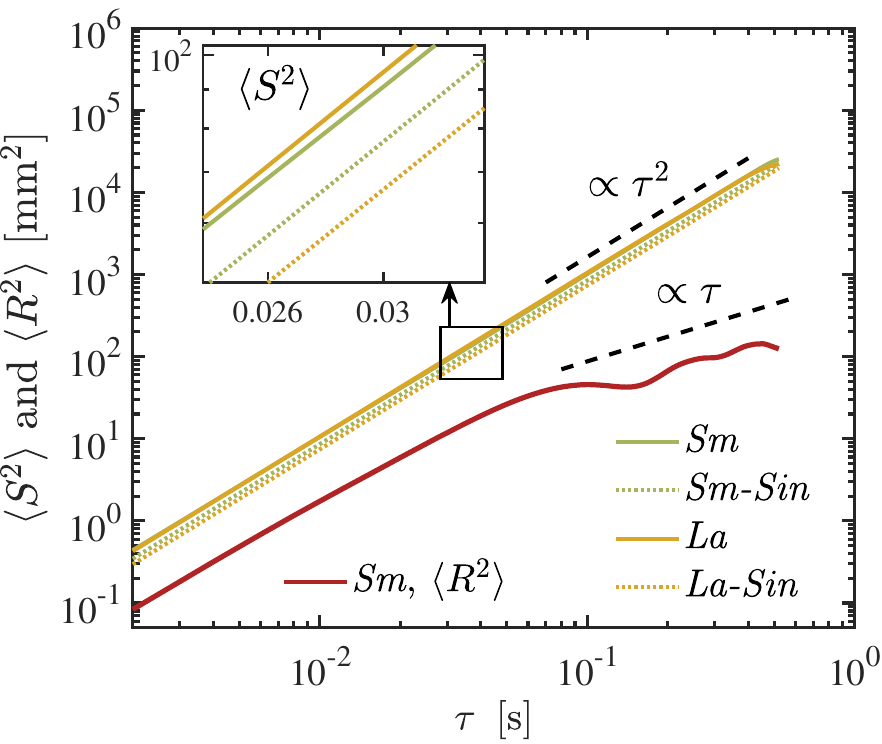}}
		\end{minipage}
		\begin{minipage}[b]{0.5\linewidth}
			\centering
			\makebox[1.2em][l]{\raisebox{-\height}{(\textit{b})}}%
			\raisebox{-\height}{\includegraphics[height=5.3cm]{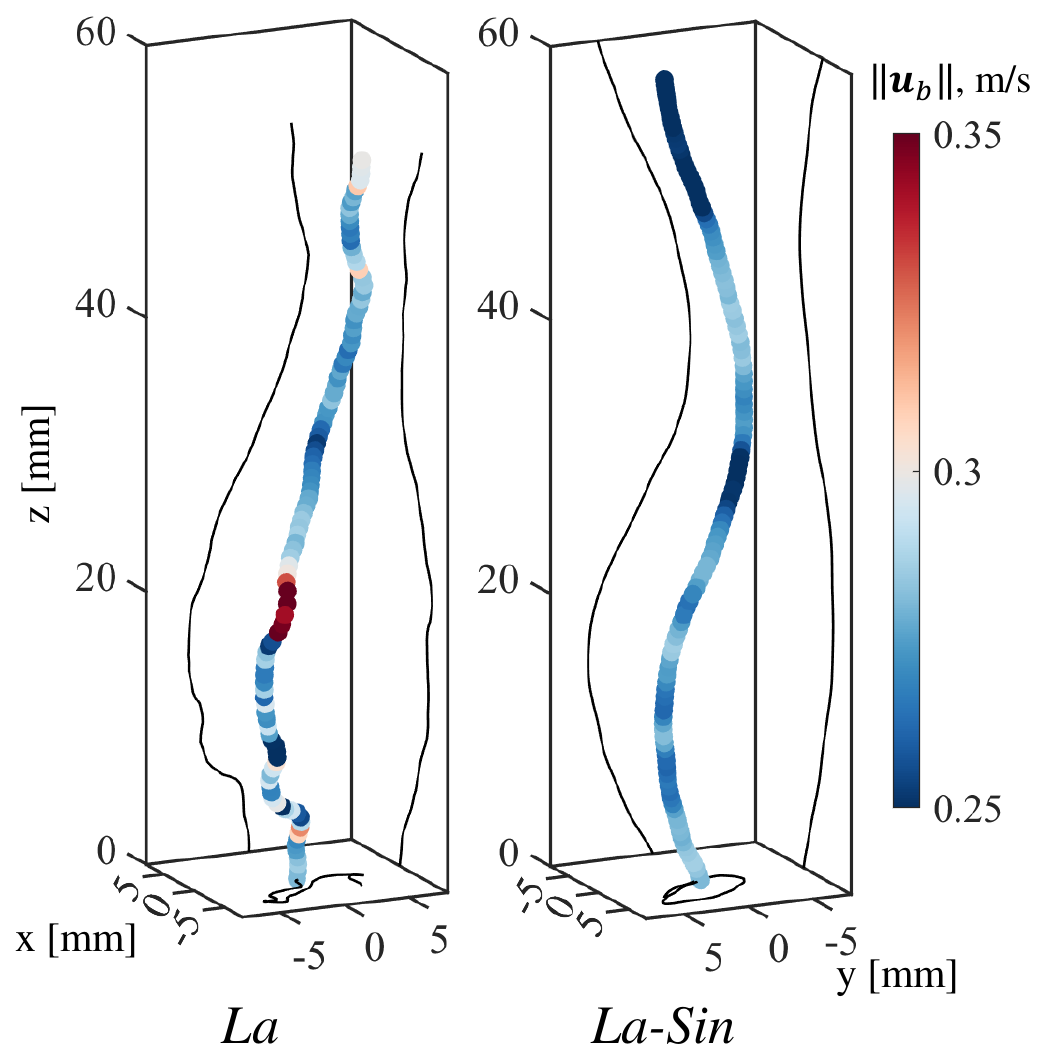}}
		\end{minipage}
	\end{minipage}	
	\caption{(\textit{a}) Mean squared path length $\langle S^2 \rangle$, together with the total mean squared displacement $\langle R^2 \rangle$ of \textit{Sm} as comparison. (\textit{b}) Snapshots of typical bubble trajectories for \textit{La} and \textit{La-Sin}, with the same tracking time. The trajectories are coloured by the magnitude of the total velocity $\left\|\boldsymbol{u}_b \right\|$.} \label{fig: path length}
\end{figure}

\subsection{Path length of bubble swarms versus single bubbles}\label{sec: 3.2}

While the MSD provides a simple metric for understanding the bubble dispersion, further information can be found by studying the path length $S(\tau)$ of the trajectories \citep{2011_Ouellette}, i.e., the total distance travelled by individual bubbles. We can approximately measure the mean-squared path length using 
\begin{equation}
\left\langle S^2(\tau) \right\rangle \approx \Big\langle \Big[ \sum_{\gamma=1}^{N_{\tau}} \left\| \mathbf{x}(t_0 + \gamma \Delta t) - \mathbf{x}(t_0 + [\gamma - 1] \Delta t) \right\|\Big]^2 \Big\rangle,
\end{equation}
where $N_\tau\equiv1+(\tau-t_0)/\Delta t \in \mathbb{Z}^+$, and $\Delta t=0.002s$ is used based on our imaging rate for tracking bubbles. For a representative case \textit{Sm}, we plot $\langle S^2\rangle$ in figure \ref{fig: path length}(\textit{a}), alongside the total MSD $\langle R^2(\tau) \rangle$ based on all 3 components of $\boldsymbol{x}$
\begin{equation} \label{eqn: MSD}
  \langle R^2(\tau) \rangle  \equiv \langle ||\Delta_{\tau}\boldsymbol{x} - \langle \Delta_{\tau}\boldsymbol{x}\rangle||^2\rangle.
\end{equation}
The other case \textit{La} shows similar behaviour and so is not shown here. For both single bubbles and bubble swarms, $\langle S^2\rangle$ grows as $\tau^2$ over almost the entire trajectory, differing from $\langle R^2\rangle$ which undergoes a transition towards a diffusive growth at long times. We see a vertical shift between $\langle S^2\rangle$ and $\langle R^2\rangle$ in the short term due to the subtraction of mean displacement by computing $\langle R^2 \rangle$ in (\ref{eqn: MSD}). (As an aside: in \cite{2011_Ouellette}, $\langle S^2\rangle\approx\langle R^2\rangle$ for the short-time ballistic regime, since in their flow there is no mean displacement for the tracer particles). The zoomed in view of $\langle S^2\rangle$ (inset of figure \ref{fig: path length}\textit{a}) shows that the path travelled by bubbles rising in swarms is longer than that for the single-bubble cases. This may seem counterintuitive at first, as it has been found that bubbles rising in turbulence have a reduced rise velocity compared to those in a quiescent liquid \citep{2017_Loisy,2021_Ruth}. The explanation for this is that the path
length is determined by $\langle\left\| \boldsymbol{u}_b \right\|\rangle = \langle\sqrt{u_b^2+v_b^2+w_b^2}\rangle$, and not the rise velocity. Indeed, the exact expression for the path length is $ S(\tau)  = \int_0^\tau \|\boldsymbol{u}_b(s)\|\,ds$, and for statistically stationary bubble tracks and large $\tau$ we have $S(\tau)\approx \tau \langle\|\boldsymbol{u}_b\|\rangle$. The measured $\langle\left\| \boldsymbol{u}_b \right\|\rangle$ is 0.315 m/s for \textit{Sm} and 0.286 m/s for \textit{Sm-Sin}, and 0.319 m/s for \textit{La} and 0.269 m/s for \textit{La-Sin}. This corresponds to a 10.2\% and 18.7\% increase in path length for \textit{Sm} and \textit{La}, respectively, compared to their single-bubble cases. The figures also indicate that the rising paths of bubbles in swarms still exhibit oscillations, and this is due to the intrinsic wake-driven dynamics dominating the low frequency behaviour of the bubble motion. However, higher frequency motion is more erratic and reflects the effect of BIT at small scales in the flow.

\subsection{PDF of fluctuating bubble velocity of bubble swarms versus single bubbles}\label{sec: 3.3}

\begin{figure}	
	\begin{minipage}[b]{1.0\linewidth}
		\begin{minipage}[b]{0.5\linewidth}
			\centering
			\makebox[1.2em][l]{\raisebox{-\height}{(\textit{a})}}%
			\raisebox{-\height}{\includegraphics[height=4.5cm]{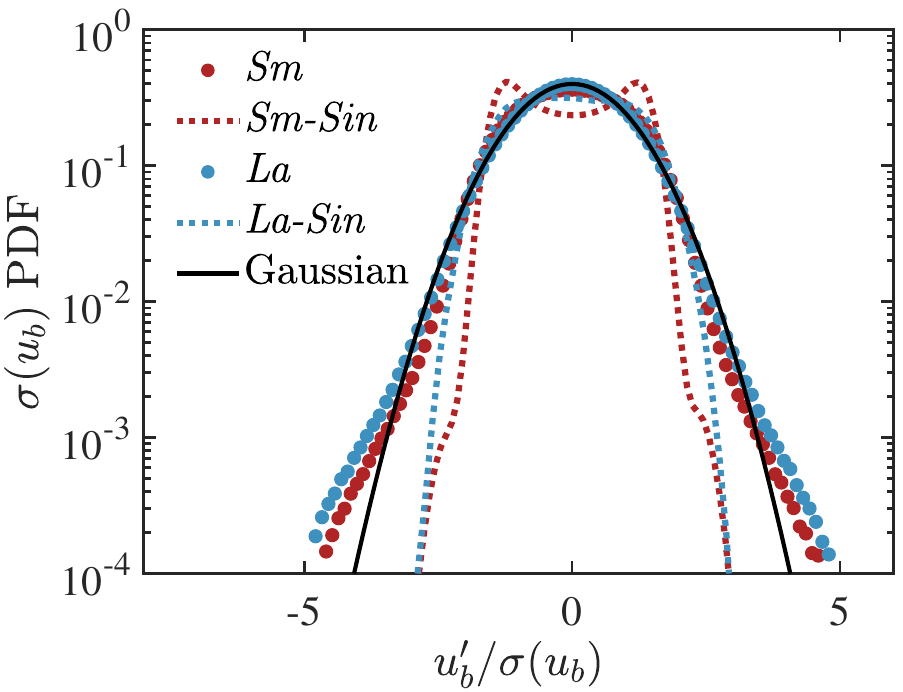}}
		\end{minipage}
		\begin{minipage}[b]{0.5\linewidth}
			\centering
			\makebox[1.2em][l]{\raisebox{-\height}{(\textit{b})}}%
			\raisebox{-\height}{\includegraphics[height=4.5cm]{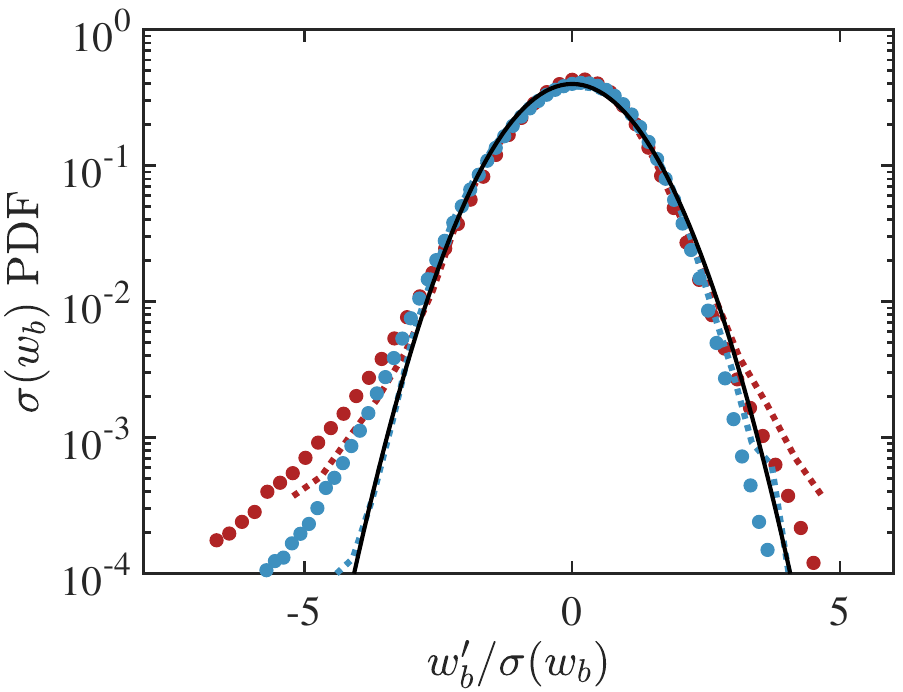}}
		\end{minipage}
	\end{minipage}	
	\caption{PDFs of the fluctuating bubble velocities in (\textit{a}) horizontal and (\textit{b}) vertical directions, normalized by their standard deviations $\sigma(u_b)$ and $\sigma(w_b)$, respectively.} \label{fig: pdf}
\end{figure}
To fully understand the bubble motion as they disperse, we now turn to consider the PDFs of the fluctuating bubble velocities, $\boldsymbol{u}'_b=\boldsymbol{u}_b-\langle\boldsymbol{u}_b\rangle$, in figure \ref{fig: pdf}. In the horizontal direction, both single bubbles and bubble swarms have symmetric and non-Gaussian PDFs. For the two single-bubble cases, while the PDF of \textit{Sm-Sin} is bi-modal, \textit{La-Sin} is uni-modal with a peak that is wide and flat compared to a Gaussian PDF. These features can also be related to the oscillations in their trajectories, with the stronger oscillations of the MSD for the case \textit{Sm-Sin} (figure \ref{fig: MSD1}\textit{a}) being associated with a clear bi-modal velocity PDF. Similar behavior was also reported by \cite{2015_Mathai} for finite-sized buoyant particles. For the corresponding swarm cases, the BIT and bubble-bubble interactions introduce additional randomness into the bubble motion, leading to Gaussian-like behavior in the central region of the PDFs. However, the PDFs have tails which are heavier than a Gaussian PDF while those for single-bubbles are lighter, indicating intermittency due to the collective motion of the bubbles in swarms. Such intermittent bubble velocities have been previously observed \citep{2022_Ma}, and may arise when a trailing bubble is caught in the wake of leading bubbles, or be due to intense hydrodynamic interactions between bubbles when they become sufficiently close.
In the vertical direction (figure \ref{fig: pdf}\textit{b}), the PDFs of \textit{Sm} and \textit{La} are negatively skewed, unlike their corresponding single-bubble cases for which the PDFs are approximately symmetric. The negative skewness in the swarm cases is consistent with our previous study \citep{2023_Ma} and the observation in \cite{2010_Martinez}. For bubble swarms, \cite{2023_Ma} argued that the PDF of bubble vertical velocity fluctuations can exhibit skewness of either sign, depending in a complex way on the background turbulence and additional factors, such as the bubble size relative to the turbulence length scales.


\subsection{Dispersion of bubble swarms versus liquid tracers}\label{sec: 3.4}

\begin{figure}	
	\begin{minipage}[b]{1.0\linewidth}
		\begin{minipage}[b]{0.5\linewidth}
			\centering
			\makebox[1.2em][l]{\raisebox{-\height}{(\textit{a})}}%
			\raisebox{-\height}{\includegraphics[width=6cm]
            {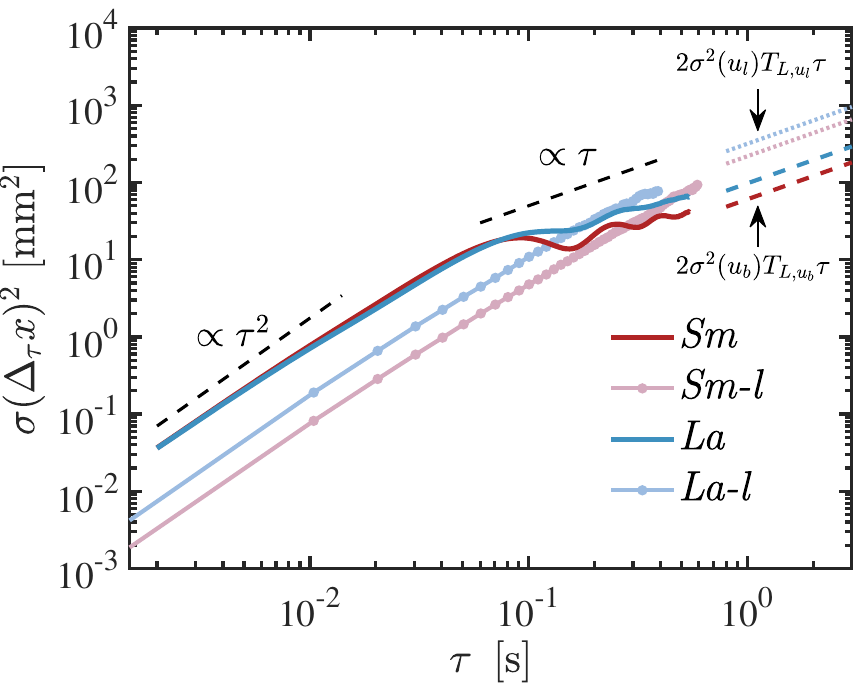}}
		\end{minipage}
		\begin{minipage}[b]{0.5\linewidth}
			\centering
			\makebox[1.2em][l]{\raisebox{-\height}{(\textit{b})}}%
			\raisebox{-\height}{\includegraphics[width=6cm]
            {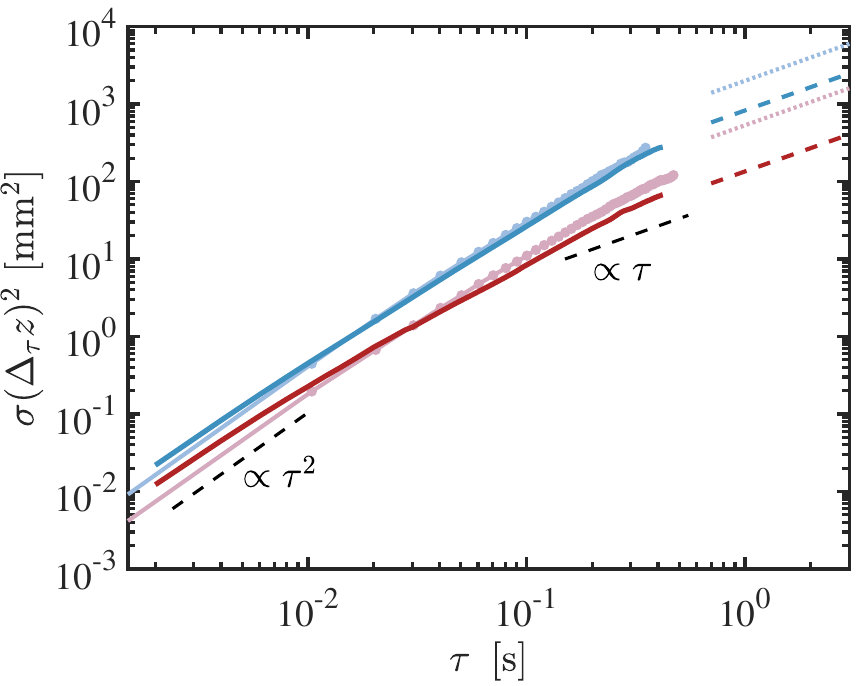}}
		\end{minipage}
	\end{minipage}	
	\begin{minipage}[b]{1.0\linewidth}
		\vspace{2mm}
		\begin{minipage}[b]{0.5\linewidth}
			\centering
			\makebox[1.6em][l]{\raisebox{-\height}{(\textit{c})}}%
			\raisebox{-\height}{\includegraphics[width=6cm]
            {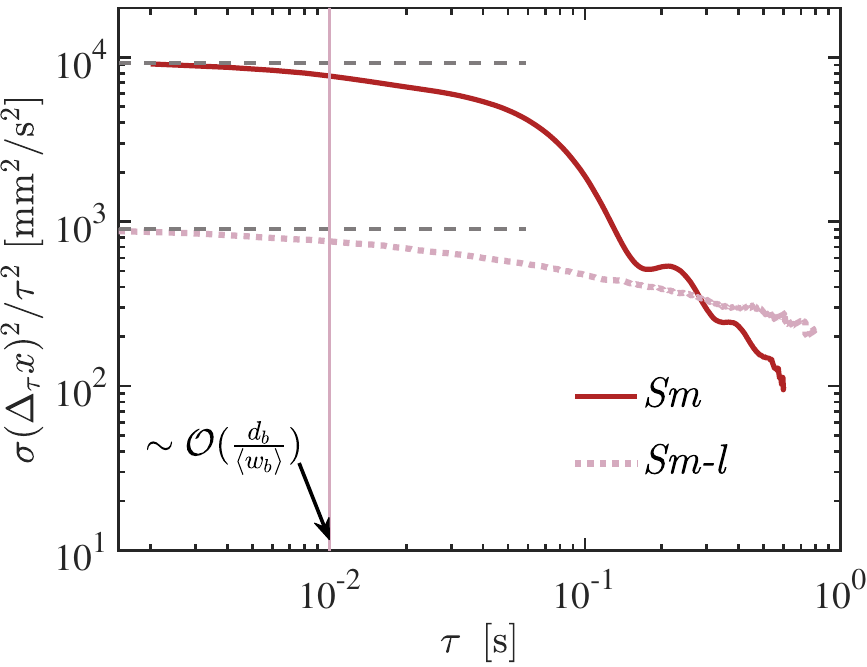}}
		\end{minipage}
		\begin{minipage}[b]{0.5\linewidth}
			\centering
			\makebox[1.6em][l]{\raisebox{-\height}{(\textit{d})}}%
			\raisebox{-\height}{\includegraphics[width=6cm]
            {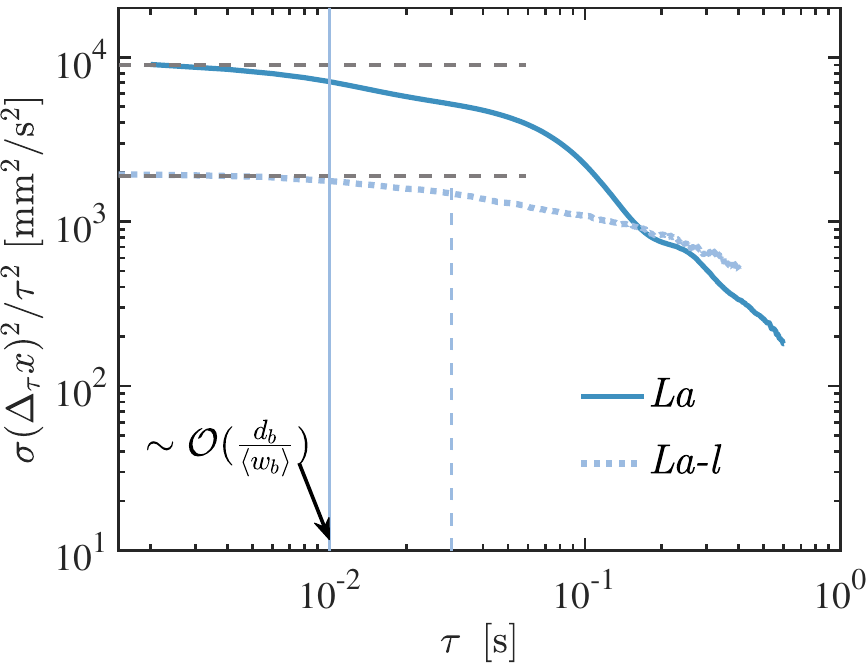}}
		\end{minipage}
	\end{minipage}	     
	\caption{MSD for bubbles and fluid tracers in the (\textit{a}) horizontal and (\textit{b}) vertical directions. Dashed lines indicate the asymptotic dispersion rate for longtime diffusive behavior. Compensated horizontal MSD for the \textit{Sm} (\textit{c}) and \textit{La} (\textit{d}) cases. The solid vertical lines indicate the onset of transition for bubbles, while the dashed vertical line marks the transition for tracers, defined as the time when the local exponential decay reaches 80\%. For \textit{Sm}, the bubble transition occurs at a similar time, so only one vertical line is shown.} \label{fig: MSD2}
\end{figure}

We finally turn our attention to the difference in dispersion between bubbles and liquid tracers in the swarm cases, and figures \ref{fig: MSD2}(\textit{a}, \textit{b}) show their horizontal and vertical MSD, respectively. At short times, the horizontal dispersion of bubbles is almost an order of magnitude faster than that of tracers, whereas the difference in their MSD in the vertical direction is smaller. This can be attributed to the relative difference between the fluctuating bubble velocity and fluctuating liquid velocity, and how this difference varies in the horizontal and vertical directions (see figure \ref{fig: exp setup}\textit{b} for the \textit{Sm} and \textit{La} cases). More specifically, the horizontal components $\sigma(u_b)$ and $\sigma(v_b)$ are significantly larger than the corresponding ones for the liquid, $\sigma(u_l)$ and $\sigma(v_l)$. In contrast, the difference between the vertical components $\sigma(w_b)$ and $\sigma(w_l)$ is much smaller, leading to more similar ballistic dispersion between the two phases (figure \ref{fig: MSD2}\textit{b}). At longer times the horizontal MSD of bubbles approaches (but does not actually attain due to the limited $\tau_{max}$, as discussed earlier) a diffusive-like regime $\propto \tau$, while the MSD for tracers does not. This is simply due to $\tau_{max}$ being too small to observe this regime for the tracers. However, the plots show the diffusive prediction for the tracers and these indicate that the diffusive regime would be reached if $\tau_{max}$ were several times larger.

With increasing $\tau$, the MSD for the tracers gradually exceeds that for the bubbles, in agreement with the results of \cite{2018_Mathai} for isolated smaller bubbles in HIT. \cite{2018_Mathai} explained this as being due to the crossing-trajectory effect, whereby bubbles can drift through turbulent eddies due to their buoyant rising motion, causing the fluid velocities along their trajectories to decorrelate more rapidly than along tracer trajectories. This mechanism is plausible for their experiments where the bubbles are rising through pre-existing background turbulence and eddies. However, its role in our experiments is less clear, because in ours the bubbles are not rising through pre-existing turbulent eddies, but are in fact generating the eddies themselves as they rise. In our case, what is clear is that the longer correlation timescale for the tracers means that the faster ballistic growth persists for longer for tracers than bubbles, enabling their dispersion to overtake that of the bubbles. The crossing trajectory effect may contribute to the reason why the correlation timescale is longer for tracers, but it could also be due to fluctuations associated with the bubble wakes causing the bubble velocities to decorrelate faster than those of tracers which do not generate fluctuating wakes due to the low Reynolds number of the tracers.

In figures \ref{fig: MSD2}(\textit{c,d}), we plot the compensated horizontal MSD for both phases and find that for both \textit{Sm} and \textit{La}, the time span of the ballistic regime is set by the wake-induced motion at a bubble timescale $O(d_b/\left< w_b \right>)$, approximately 0.01 s. However, the time span for the tracers is different for the \textit{Sm} and \textit{La} cases. For the \textit{La} case, the transition away from the ballistic regime occurs considerably later for the tracers than for the bubbles for the \textit{La} case. This confirms the finding of \cite{2018_Mathai} that the transition for bubbles occurs much earlier than for tracer particles due to their oscillatory wake-driven dynamics. By contrast, for the \textit{Sm} case the transition for bubbles and tracers occurs at the same time (see figure \ref{fig: MSD2}\textit{c}). The explanation is that for the \textit{Sm} case where $\alpha$ is much smaller, there are relatively few regions in the liquid where turbulence is generated by the bubbles, such that turbulence in the flow is very patchy and intermittent. In this case, the tracer dispersion will be highly dependent upon the bubble parameters. For the \textit{La} case with higher $\alpha$ and larger $d_b$, the bubble wakes are longer, wider, and more numerous. As a result, the wake regions are more space filling and wake-wake interactions are more frequent leading to a flow that is more homogeneous and well mixed (see \cite{2025_Ma}). In this respect, the flow in the \textit{La} case is more similar to that in \cite{2018_Mathai} where there is background turbulence generated by an active grid. This explains the qualitative agreement between the \textit{La} case and the results of \cite{2018_Mathai} regarding the timing of the departure from the ballistic regime for both tracers and bubbles. However, quantitatively, \cite{2018_Mathai} observe the bubbles departing from the ballistic regime ten times faster than the tracers, whereas in our case it is three times faster. This difference is due to most of the turbulence in their flow being generated by the active grid, whereas in ours it is purely BIT which is more dependent on the bubble parameters.

\section{Conclusions}

We used experiments equipped with 3D-LBT and 3D-LPT to explore the Taylor dispersion of bubble swarms rising in an initially quiescent water. The bubbles considered are large and relatively dense, and their wakes generate significant turbulence and bubble-bubble interactions -- a complementary flow condition to that in the previous work by \cite{2018_Mathai}. To better understand their Taylor dispersion we compared: (i) bubble swarms with single bubbles; (ii) bubble swarms with fluid tracers. 

For (i), we observe that at long times the horizontal MSD exhibits oscillations for both the bubble swarms and single-bubble cases. These oscillations are due to wake-induced motion of the bubbles and are, however, strongly damped for the swarm cases compared to the single-bubble cases, primarily due to BIT and bubble-bubble interactions. Due to the damping, the swarm cases approach a diffusive regime, while the single-bubble cases do not. The vertical MSD in bubble swarms is significantly larger than that for the single-bubble cases due to the much higher vertical fluctuating bubble velocity in the swarm cases. We also compute the path length of the rising bubbles and find that it is up to $18.7\%$ larger in swarms than for single bubbles due to BIT which contorts the trajectories in the swarms.

For (ii), we find a faster dispersion rate in the ballistic regime for bubble swarms than for tracers due to their larger velocities. At longer times, the dispersion of the bubbles is overtaken by the tracers as the faster ballistic regime persists for longer for the tracers. The time to transition away from the ballistic regime is of the order of bubble timescale for smaller and more dilute swarm cases, while the results for larger bubbles with higher gas void fraction reveal an earlier transition for the bubbles compared to the tracers.

\section*{Acknowledgements}
We thank the financial support from German Research Foundation under Grant 555712173.

\section*{Declaration of Interests}
The authors report no conflict of interest.

\bibliographystyle{jfm}

\bibliography{BIT_exp_JFM}

\begin{thebibliography}{23}
\expandafter\ifx\csname natexlab\endcsname\relax\def\natexlab#1{#1}\fi
\def\au#1{#1} \def\ed#1{#1} \def\yr#1{#1}\def\at#1{#1}\def\jt#1{\textit{#1}}
  \def\bt#1{#1}\def\bvol#1{\textbf{#1}} \def\vol#1{#1} \def\pg#1{#1}
  \def\publ#1{#1}\def\arxiv#1{#1}\def\org#1{#1}\def\st#1{\textit{#1}}

\bibitem[Bourgoin {\em et~al.\/}(2020)Bourgoin, Kervil, Cottin-Bizonne, Raynal,
  Volk \& Ybert]{2020_Bourgoin}
{\sc \au{Bourgoin, M.}, \au{Kervil, R.}, \au{Cottin-Bizonne, C.}, \au{Raynal,
  F.}, \au{Volk, R.} \& \au{Ybert, C.}} \yr{2020}  \at{Kolmogorovian active
  turbulence of a sparse assembly of interacting marangoni surfers}.  \jt{Phys.
  Rev. X}  \bvol{10}~(2),  \pg{021065}.

\bibitem[Bourgoin \& Xu(2014)]{2014_Bourgoin}
{\sc \au{Bourgoin, M.} \& \au{Xu, H.}} \yr{2014}  \at{Focus on dynamics of
  particles in turbulence}.  \jt{New J. Phys.}  \bvol{16}~(8),  \pg{085010}.

\bibitem[Brandt \& Coletti(2022)]{2022_Brandt}
{\sc \au{Brandt, L.} \& \au{Coletti, F.}} \yr{2022}  \at{Particle-laden
  turbulence: progress and perspectives}.  \jt{Annu. Rev. Fluid Mech.}
  \bvol{54},  \pg{159--189}.

\bibitem[Hessenkemper {\em et~al.\/}(2024)Hessenkemper, Wang, Lucas, Tan, Ni \&
  Ma]{2024_Hessenkemper}
{\sc \au{Hessenkemper, H.}, \au{Wang, L.}, \au{Lucas, D.}, \au{Tan, S.},
  \au{Ni, R.} \& \au{Ma, T.}} \yr{2024}  \at{3d detection and tracking of
  deformable bubbles in swarms with the aid of deep learning models}.  \jt{Int.
  J. Multiphase Flow}  \bvol{179},  \pg{104932}.

\bibitem[Innocenti {\em et~al.\/}(2021)Innocenti, Jaccod, Popinet \&
  Chibbaro]{2021_Innocenti}
{\sc \au{Innocenti, A.}, \au{Jaccod, A.}, \au{Popinet, S.} \& \au{Chibbaro,
  S.}} \yr{2021}  \at{Direct numerical simulation of bubble-induced
  turbulence}.  \jt{J. Fluid Mech.}  \bvol{918}.

\bibitem[Legendre \& Zenit(2025)]{2025_Legendre}
{\sc \au{Legendre, D.} \& \au{Zenit, R.}} \yr{2025}  \at{Gas bubble dynamics}.
  \jt{RMP, arXiv preprint arXiv:2501.02988} .

\bibitem[Lohse(2018)]{2018_Lohse}
{\sc \au{Lohse, D.}} \yr{2018}  \at{Bubble puzzles: From fundamentals to
  applications}.  \jt{Phys. Rev. Fluids}  \bvol{3}~(11),  \pg{110504}.

\bibitem[Loisy \& Naso(2017)]{2017_Loisy}
{\sc \au{Loisy, A.} \& \au{Naso, A.}} \yr{2017}  \at{Interaction between a
  large buoyant bubble and turbulence}.  \jt{Phys. Rev. Fluids}  \bvol{2}~(1),
  \pg{014606}.

\bibitem[Lu \& Tryggvason(2013)]{2013_Lu}
{\sc \au{Lu, J.} \& \au{Tryggvason, G.}} \yr{2013}  \at{Dynamics of nearly
  spherical bubbles in a turbulent channel upflow}.  \jt{J. Fluid Mech.}
  \bvol{732},  \pg{166--189}.

\bibitem[Ma {\em et~al.\/}(2022)Ma, Hessenkemper, Lucas \& Bragg]{2022_Ma}
{\sc \au{Ma, T.}, \au{Hessenkemper, H.}, \au{Lucas, D.} \& \au{Bragg, A.~D.}}
  \yr{2022}  \at{An experimental study on the multiscale properties of
  turbulence in bubble-laden flows}.  \jt{J. Fluid Mech.}  \bvol{936},
  \pg{A42}.

\bibitem[Ma {\em et~al.\/}(2023)Ma, Hessenkemper, Lucas \& Bragg]{2023_Ma}
{\sc \au{Ma, T.}, \au{Hessenkemper, H.}, \au{Lucas, D.} \& \au{Bragg, A.~D.}}
  \yr{2023}  \at{Fate of bubble clusters rising in a quiescent liquid}.  \jt{J.
  Fluid Mech.}  \bvol{973},  \pg{A15}.

\bibitem[Ma {\em et~al.\/}(2025)Ma, Tan, Ni, Hessenkemper \& Bragg]{2025_Ma}
{\sc \au{Ma, T.}, \au{Tan, S.}, \au{Ni, R.}, \au{Hessenkemper, H.} \&
  \au{Bragg, A.~D.}} \yr{2025}  \at{Kolmogorov scaling in bubble-induced
  turbulence}.  \jt{Phys. Rev. Lett., (accept)} .

\bibitem[Mart\'{i}nez {\em et~al.\/}(2010)Mart\'{i}nez, Chehata, van Gils, Sun
  \& Lohse]{2010_Martinez}
{\sc \au{Mart\'{i}nez, M.~J.}, \au{Chehata, G.~D.}, \au{van Gils, D.}, \au{Sun,
  C.} \& \au{Lohse, D.}} \yr{2010}  \at{On bubble clustering and energy spectra
  in pseudo-turbulence}.  \jt{J. Fluid Mech.}  \bvol{650},  \pg{287--306}.

\bibitem[Mathai {\em et~al.\/}(2018)Mathai, Huisman, Sun, Lohse \&
  Bourgoin]{2018_Mathai}
{\sc \au{Mathai, V.}, \au{Huisman, S.~G.}, \au{Sun, C.}, \au{Lohse, D.} \&
  \au{Bourgoin, M.}} \yr{2018}  \at{Dispersion of air bubbles in isotropic
  turbulence}.  \jt{Phys. Rev. Lett.}  \bvol{121}~(5),  \pg{054501}.

\bibitem[Mathai {\em et~al.\/}(2020)Mathai, Lohse \& Sun]{2020_Mathai}
{\sc \au{Mathai, V.}, \au{Lohse, D.} \& \au{Sun, C.}} \yr{2020}  \at{Bubbly and
  buoyant particle--laden turbulent flows}.  \jt{Annu. Rev. Condens. Matter
  Phys. 11}  \bvol{11},  \pg{529--559}.

\bibitem[Mathai {\em et~al.\/}(2015)Mathai, Prakash, Brons, Sun \&
  Lohse]{2015_Mathai}
{\sc \au{Mathai, V.}, \au{Prakash, V.~N.}, \au{Brons, J.}, \au{Sun, C.} \&
  \au{Lohse, D.}} \yr{2015}  \at{Wake-driven dynamics of finite-sized buoyant
  spheres in turbulence}.  \jt{Phys. Rev. Lett.}  \bvol{115}~(12),
  \pg{124501}.

\bibitem[Ouellette {\em et~al.\/}(2011)Ouellette, Bodenschatz \&
  Xu]{2011_Ouellette}
{\sc \au{Ouellette, N.~T.}, \au{Bodenschatz, E.} \& \au{Xu, H.}} \yr{2011}
  \at{Path lengths in turbulence}.  \jt{J. Stat. Phys.}  \bvol{145},
  \pg{93--101}.

\bibitem[Riboux {\em et~al.\/}(2010)Riboux, Risso \& Legendre]{2010_Riboux}
{\sc \au{Riboux, G.}, \au{Risso, F.} \& \au{Legendre, D.}} \yr{2010}
  \at{Experimental characterization of the agitation generated by bubbles
  rising at high {R}eynolds number}.  \jt{J. Fluid Mech.}  \bvol{643},
  \pg{509--539}.

\bibitem[Ruth {\em et~al.\/}(2021)Ruth, Vernet, Perrard \& Deike]{2021_Ruth}
{\sc \au{Ruth, D.~J.}, \au{Vernet, M.}, \au{Perrard, S.} \& \au{Deike, L.}}
  \yr{2021}  \at{The effect of nonlinear drag on the rise velocity of bubbles
  in turbulence}.  \jt{J. Fluid Mech.}  \bvol{924},  \pg{A2}.

\bibitem[Schanz {\em et~al.\/}(2016)Schanz, Gesemann \&
  Schr{\"o}der]{2016_Schanz}
{\sc \au{Schanz, D.}, \au{Gesemann, S.} \& \au{Schr{\"o}der, A.}} \yr{2016}
  \at{Shake-the-box: Lagrangian particle tracking at high particle image
  densities}.  \jt{Exp. Fluids}  \bvol{57},  \pg{1--27}.

\bibitem[Tan {\em et~al.\/}(2020)Tan, Salibindla, Masuk \& Ni]{2020_Tan}
{\sc \au{Tan, S.}, \au{Salibindla, A.}, \au{Masuk, A. U.~M.} \& \au{Ni, R.}}
  \yr{2020}  \at{Introducing openlpt: new method of removing ghost particles
  and high-concentration particle shadow tracking}.  \jt{Exp. Fluids}
  \bvol{61},  \pg{1--16}.

\bibitem[Taylor(1922)]{1922_Taylor}
{\sc \au{Taylor, G.~I.}} \yr{1922}  \at{Diffusion by continuous movements}.
  \jt{Proc. Lond. Math. Soc.}  \bvol{2}~(1),  \pg{196--212}.

\bibitem[Toschi \& Bodenschatz(2009)]{2009_Toschi}
{\sc \au{Toschi, F.} \& \au{Bodenschatz, E.}} \yr{2009}  \at{Lagrangian
  properties of particles in turbulence}.  \jt{Annu. Rev. Fluid Mech.}
  \bvol{41}~(1),  \pg{375--404}.

\end{thebibliography}

\end{document}